\documentclass[prb,twocolumn,aps,amssymb,amsmath,superscriptaddress,showpacs]{revtex4-1}

\usepackage{bm}
\usepackage{float}
\usepackage{graphicx}
\usepackage{color}
\usepackage{mathtools}
\usepackage{amsfonts}
\usepackage{textcomp} 
\usepackage{microtype} 

\usepackage{hyperref}
\hypersetup{colorlinks=true}
\usepackage[all]{hypcap} 

\usepackage{csquotes}

\begin{document}

\title{Pairing Mechanism in Hund’s Metal Superconductors and the Universality of the
Superconducting Gap to Critical Temperature Ratio}

\author{Tsung-Han Lee}
\affiliation{Physics and Astronomy Department, Rutgers University, Piscataway, New Jersey 08854, USA}
\author{Andrey Chubukov}
\affiliation{School of Physics and Astronomy, University of Minnesota, Minneapolis, Minnesota 55455, USA}
\author{Hu Miao}
\affiliation{Brookhaven National Laboratory, Upton, New York 11973, USA}
\author{Gabriel Kotliar}
\affiliation{Physics and Astronomy Department, Rutgers University, Piscataway, NJ 08854, USA}
\affiliation{Brookhaven National Laboratory, Upton, New York 11973, USA}

\begin{abstract}
We analyze  a simple model containing the  physical ingredients  of a  Hund's metal,  the   local spin fluctuations
with   power-law correlators, $(\Omega_0/|\Omega|)^\gamma$, with  $\gamma$  greater than one,  interacting with electronic quasiparticles.
While the     critical temperature and the  gap change significantly  with varying parameters,  the   $2\Delta_{\max}/k_BT_c$  remains close to twice the BCS value in agreement with experimental observations  in the   iron-based superconductors (FeSC).
 \end{abstract}

\date{\today}

\maketitle

{\bf \it Introduction}~~~
The discovery of superconductivity in iron-based materials \cite{Hosono}
 opened a
new area of research in the field of superconducting materials. There are
by now many families of materials which are based on tetrahedrally
coordinated irons to pnictides and chalcogenides, with different
separating layers in between. For a review, see Refs. \onlinecite{IronbaseReview1} and \onlinecite{IronbaseReview2}.
Photoemission studies have shown that some  of these compounds have bands
which are well described by the standard density functional theory with
small renormalizations, while in others, the  mass renormalizations are
larger than ten. Hence, it is agreed upon that in this class of
compounds the strength of the correlation varies substantially. On the
other hand, there is no agreement on the type of correlation, which is
attributed  to Mott \cite{Si_Abrahams_Mott}  or Hund's  physics \cite{Werner_power_law,HauleHunds,Georges_Hunds_review}.

Thinking of the iron pnictides as Hund's metals presents a
scenario, in which the physics governing the behavior of different materials is the same, but
 the
 correlations
  are  sensitive to  the
filling of the shell and the  height of the pnictogen or chalcogen ligand
\cite{Yin_ironbase_nature}. A deeper understanding of Hund's metal
physics  shows that  the normal state above the superconducting
transition  has  a broad intermediate region of temperatures
characterized by orbital spin separation, whereby  the spin excitations
are quasi-atomic-like, while the orbital excitations are fully itinerant \cite{Yin_susceptibility,Aron_spin_orbital,NRG_susceptibility}.
{ Thus, the Hund's metal behavior exists in a temperature range below the Kondo scale of the orbital degrees of freedom $T_K^{\text{orb}}$ and above the Kondo scale of the spin degrees of freedom $T_{K}^{\text{sp}}$,
below which the Fermi liquid holds \cite{NRG_susceptibility}.}

   Whether a Hund's metal becomes a superconductor at high
temperatures  is expected to depend  on many microscopic details such as
the shape of the Fermi surfaces of the electrons, the dispersion of the
spin excitations, and how  they are coupled to electrons \cite{Dai_Yin_1, Dai_Yin_2}. Hence the superconducting  critical temperature
 is not a universal
quantity, much like 
a  coherence-incoherence crossover
 where a
  Fermi liquid emerges from 
   a Hund's metal state.

In this Letter, we  point out
 a universal aspect of
  superconductivity,
   which emerges from  a Hund's metal state at higher
temperatures
-- we argue that while $T_c$ and the
  maximum value of the superconducting gap at $T=0$, $\Delta_{\max}$, are material-dependent, their ratio $2\Delta_{\max}/T_c$ is material-independent universal number.
 We show that this is the case if the pairing in a Hund's metal is mediated by quasilocal spin excitations.
 As the
normal state of a Hund's metal,
involves incomplete screening, it is characterized by a power-law behavior of all
the physical quantities \cite{Yin_susceptibility,Aron_spin_orbital,NRG_susceptibility,Werner_power_law}.
In particular,
the susceptibility of local spin fluctuations has a power law dependence
above
 a  characteristic Kondo scale. While  an analytic theory of
  such power-law behavior  is not yet available,
   the numerical studies
 and physical considerations clearly indicate
 that the spin susceptibility follows  $\chi(\Omega) \propto  1/|\Omega|^{\gamma}$ with $\gamma>1$ \cite{NRG_susceptibility}. Here we show that,  when
  such $\chi (\Omega)$ mediates superconductivity
emerging from the Hund's metal state,
 the ratio  $2 \Delta_{\max}/k_BT_c$ is a universal, $\gamma-$dependent number, which for $\gamma >1$
  is  substantially
larger than the BCS value.  Universality here means that this number does not depend on the strength of the coupling to magnetic fluctuations, while
 $T_c$ and
 $\Delta_{\max}$
 vary strongly with the strength of fermion-boson  coupling.

These results are in agreement with  the conclusions of
recent experiments on FeSC  which addressed this question from an experimental perspective.
 By measuring the gap and the critical
temperature in LiFeAs and FeTe$_{0.55}$Se$_{0.45}$, Miao {\em et al.} established~
\cite{Miao_gap_Tc}  that in both systems
 $2\Delta_{\max}/k_BT_c\sim 7.2$,
  despite the fact that the electronic structures are different.
  The previous study on
the spin resonance also found a universal ratio, $\Omega_{\text{res}}/k_BT_c$ \cite{FeSC_spin-resonance-ratio}.
This last observation is consistent with the universality of $2\Delta_{\max}/k_BT_c$ if $\Omega_{\text{res}}$ scales with $\Delta_{\max}$ as numerous studies of spin resonance suggested~\cite{Eschrig}.

 Hund's metals are not confined to the iron-based superconductors and are
in fact very common. {Sr$_2$RuO$_4$}
 is a prime example
of Hund's metal \cite{Mravlje_thermal_power_PRL,Mravlje_Ru_dmft,Dang_Ru_PRL,Dang_Ru_PRB, Xiaoyu_Ru_PRL}.
{ However, for Sr$_2$RuO$_4$, its $T_{K}^{\text{sp}}$ is much higher than its superconducting
temperature \cite{Mravlje_Ru_dmft}. Consequently, the normal state above the superconducting transition is already a Fermi liquid instead of a Hund's metal as in FeSCs. Therefore, our theory does not apply there.}

{\bf \it Model}~~~
To describe superconductivity in Hund's metals we use
 the $\gamma$-model,
which was introduced in the context of
superconductivity near a quantum
critical point~
\cite{Abanov_review_normal_phase,spin-fluct-d-wave-book,Abanov_QC2,acn,Moon_gamma_model,Wang_gamma_model,She_gamma_model,Moon_SDW_Eliashberg,max_last,raghu}.
 Namely, we assume that interaction between fermions is mediated by a local spin susceptibility $\chi (\Omega) \propto 1/|\Omega|^\gamma$ \cite{supplement}.
This interaction simultaneously gives rise to pairing and frequency-dependent fermionic self-energy $\Sigma (\omega)$.
The $\gamma$-model  ignores  many of the complications of a realistic description
of the iron pnictide superconductors: a) its multiband and multiorbital
nature, b) multiple Fermi surfaces,  c)
 orbital-induced gap variation along the Fermi surfaces and the variation of the phase 
  and magnitude
  of
 a  superconducting
  order parameter between different Fermi surfaces,
 d)  fine features in the
   dynamical structure factor of spin fluctuations (see Refs. \onlinecite{review_FE1,review_FE2,review_FE3,review_FE4,review_FE5,Georges_Hunds_review} for recent reviews).
  It retains, however, two essential features, the superlinear
divergence of the local spin susceptibility  at intermediate frequencies,
and the coupling of quasilocalized spins to fermionic quasiparticles.  We argue
that this is the essential ingredient to understand the results of Ref. [\onlinecite{Miao_gap_Tc}]
 that
 a)  the  ratio  of $2\Delta_{\max} / k_B T_c $  is much larger than in BCS theory and
 b) it does not vary between different materials, as opposed to $T_c$ and $\Delta_{\max}$, both of which are
  material dependent.

The expressions for $T_c$ and $\Delta_{\max}$ in the $\gamma$-model  are obtained  by solving the set of Eliashberg equations~\cite{Eliashberg,Abanov_QC2,Moon_gamma_model,Wang_gamma_model,raghu,supplement} for the pairing vertex $\Phi (\omega_n)$ and
 fermionic self-energy $\Sigma (\omega_n)$ 
  with a power-law form of the interaction:
\begin{align}
\label{eq:Eeq1}
\Sigma(\omega_n)=\pi T\sum_{\omega_m}\lambda(\omega_m-\omega_n)\frac{\omega_m+\Sigma(\omega_m)}{\sqrt{\big[\omega_m+\Sigma(\omega_m)\big]^2+\Phi^2(\omega_m)}},\\
\label{eq:Eeq2}
\Phi(\omega_n)=\pi T\sum_{\omega_m}\lambda(\omega_m-\omega_n)\frac{\Phi(\omega_m)}{\sqrt{\big[\omega_m+\Sigma(\omega_m)\big]^2+\Phi^2(\omega_m)}},
\end{align}
 where
\begin{eqnarray}
\lambda(\Omega)=\Big(\frac{\Omega_0}{|\Omega|}\Big)^\gamma,
\label{eq:lambda}
\end{eqnarray}
 and $\Omega_0$ determines the strength of fermion-boson coupling.
The two Eliashberg  equations can be partly factorized by introducing 
the pairing gap $\Delta(\omega_n)=\Phi(\omega_n) \omega_n/\big[\omega_n + \Sigma (\omega_n)\big]$ instead of $\Phi (\omega)$.   With this substitution,
 the self-energy $\Sigma (\omega_n)$ drops from the equation for $\Delta (\omega)$. We have
\begin{align}
\Delta(\omega_n)=\pi T \sum_{\omega_m}\frac{\lambda(\omega_m-\omega_n)}{\sqrt{\omega_m^2+\Delta^2(\omega_m)}}\big(\Delta(\omega_m)-\Delta(\omega_n)\frac{\omega_m}{\omega_n} \big).
\label{eq:Delta}
\end{align}
{ The $\lambda (\Omega)$ diverges at $\Omega=0$, when $\omega_n=\omega_m$. However, the term in the bracket in the rhs of Eq. \ref{eq:Delta} becomes zero at $\omega_n=\omega_m$, which cancels out the divergence. Hence, Eq. \ref{eq:Delta} is free from singularities at any finite $T$.}
The equation on $\Sigma(\omega)$ does depend on $\Delta (\Omega_n)$:
\begin{align}
\Sigma(\omega_n)=\pi T \sum_{\omega_m}\lambda(\omega_m-\omega_n)\frac{\omega_m}{\sqrt{\omega_m^2+\Delta^2(\omega_m)}}.
\label{eq:Z}
\end{align}

\vspace{0.cm}
\begin{figure}[h]
\includegraphics[width=8.5cm]{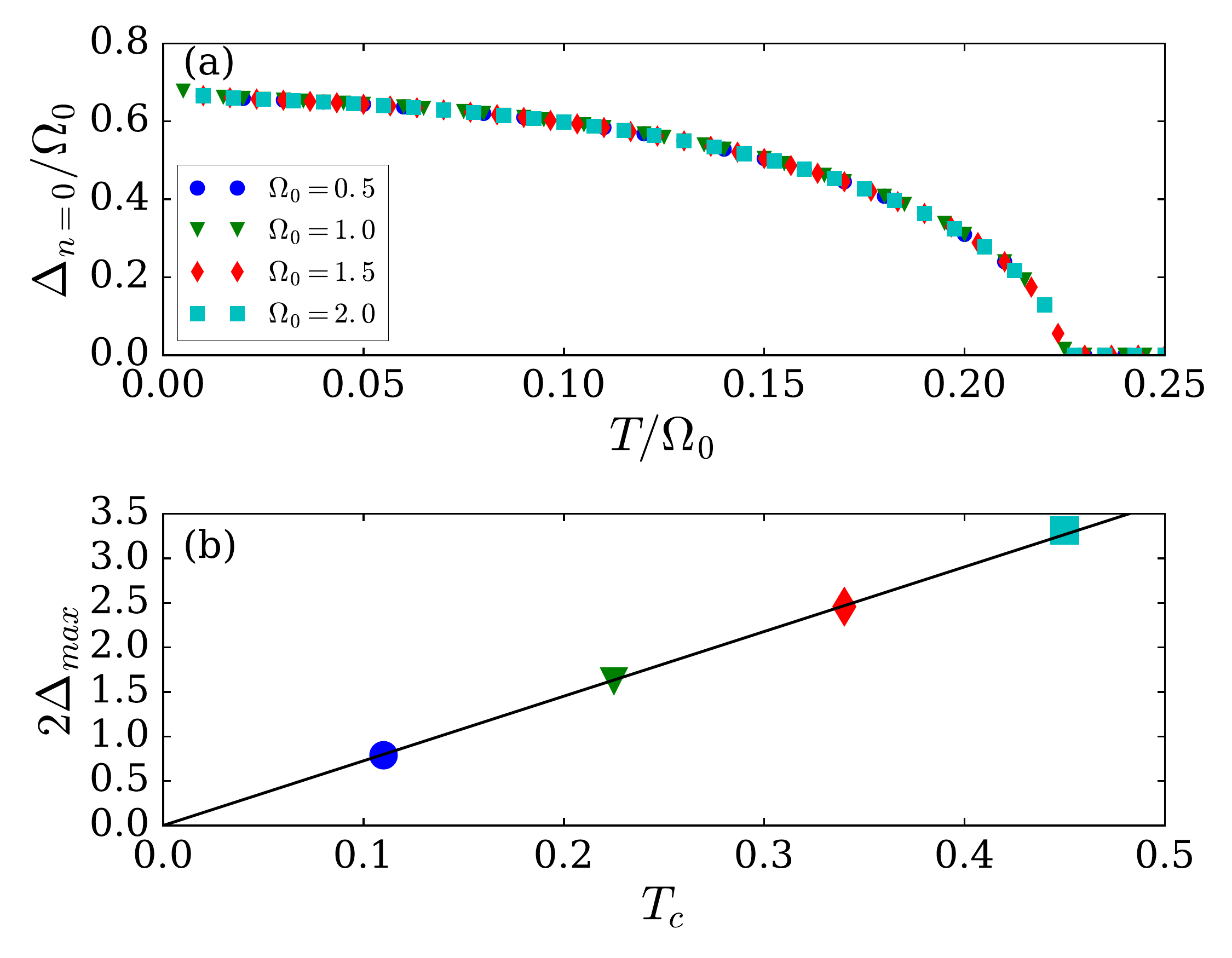}
\caption{(Color online) (a) The pairing gap at the first Matsubara frequency
 $\Delta_{n=0}$ as a function of temperature $T$ for different pairing amplitudes $\Omega_0$ at $\gamma=1.2$. (b) The maximum gap $\Delta_{\max}$ and critical temperature $T_c$ for various paring amplitudes $\Omega_0$ at $\gamma=1.2$. The black solid line is the linear fit to the slope, $2\Delta_{\max}/T_c=7.2$, corresponding to the experimental value observed in FeSC.}
\label{fig1}
\end{figure}

\noindent Because of semifactorization, one has to solve first Eq. \ref{eq:Delta} for $\Delta (\omega_n)$, substitute the result into Eq. \ref{eq:Z} and obtain $\Sigma (\omega_n)$.

 In quantum-critical theories, $\gamma=2$ corresponds to the
strong coupling limit of electron-phonon interaction \cite{strong_BCS}, $\gamma=1/2$ describes pairing by  antiferromagnetic spin fluctuations in 2D\cite{Abanov_review_normal_phase,Abanov_QC2,acn}, {$\gamma =1/3$} describes pairing by a gauge field and ferromagnetic spin fluctuations in 2D
~\cite{nick_b,Chubukov_2003_ferro,Klein_Chubukov_2018_ferro,Altshuler_13,Millis_ferro,max_last}, and $\gamma=0+$ describes color superconductivity and pairing in 3D~\cite{Son_colorSC,Chubukov_colorSC}. The models with varying $\gamma <1$ have also been analyzed~\cite{Moon_SDW_Eliashberg,Moon_gamma_model,Wang_gamma_model,raghu}.
  Here we use the fact that 
  in a wide range of frequencies a Hund's metal is also characterized by a  local
susceptibility, $\chi(\Omega)\propto 1/|\Omega|^{\gamma}$, with $\gamma$ greater than one \cite{supplement}, and explore the consequences of such a model on the $2\Delta_{\max} / T_c $ ratio by numerically and analytically solving Eq. \ref{eq:lambda} and Eq. \ref{eq:Delta}. We  obtain $\Delta (\omega_m)$ on the Matsubara axis and convert it onto real axis by analytical continuation. We define $\Delta_{\max}$ at $T=0.005\Omega_0$ as the frequency at which the density of states $N(\omega) \propto Im\big[ \omega/(\Delta^2 (\omega) - \omega^2) \big]$ jumps to a finite value, i.e., set $\Delta_{\max} = \Delta (\omega = \Delta_{\max})$.

{\bf \it The results}~~~
Figure \ref{fig1}(a) shows our results of the pairing gap at the first Matsubara frequency, $\Delta_{n=0}$, as a function of temperature $T$ for
 a given $\gamma =1.2$.  We see that $\Delta_{n=0}$, measured in units of the interaction strength $\Omega_0$, is a universal function of $T/\Omega_0$ (i.e., the functional form
  does not depend on $\Omega_0$). This can be seen directly from  Eq. \ref{eq:Delta} by simultaneously rescaling $\Delta (\omega_n)$ and Matsubara frequencies $\omega_{n,m}$ by $\Omega_0$.
 For this particular $\gamma$ we obtained $\Delta_{\max}=0.69\Omega_0
  \approx \Delta (\pm \pi T)$
  and $T_c=0.19\Omega_0$.
 The ratio $2\Delta_{\max}/T_c = 7.2$ is the universal number, independent of $\Omega_0$, as we explicitly show in Fig. \ref{fig1}(b).
     This universality is indeed the consequence of the fact that  $\Omega_0$ is the only energy scale in the problem.
    For a generic $\gamma$, we expect

\vspace{0.cm}
\begin{figure}[h]
\includegraphics[width=8.5cm]{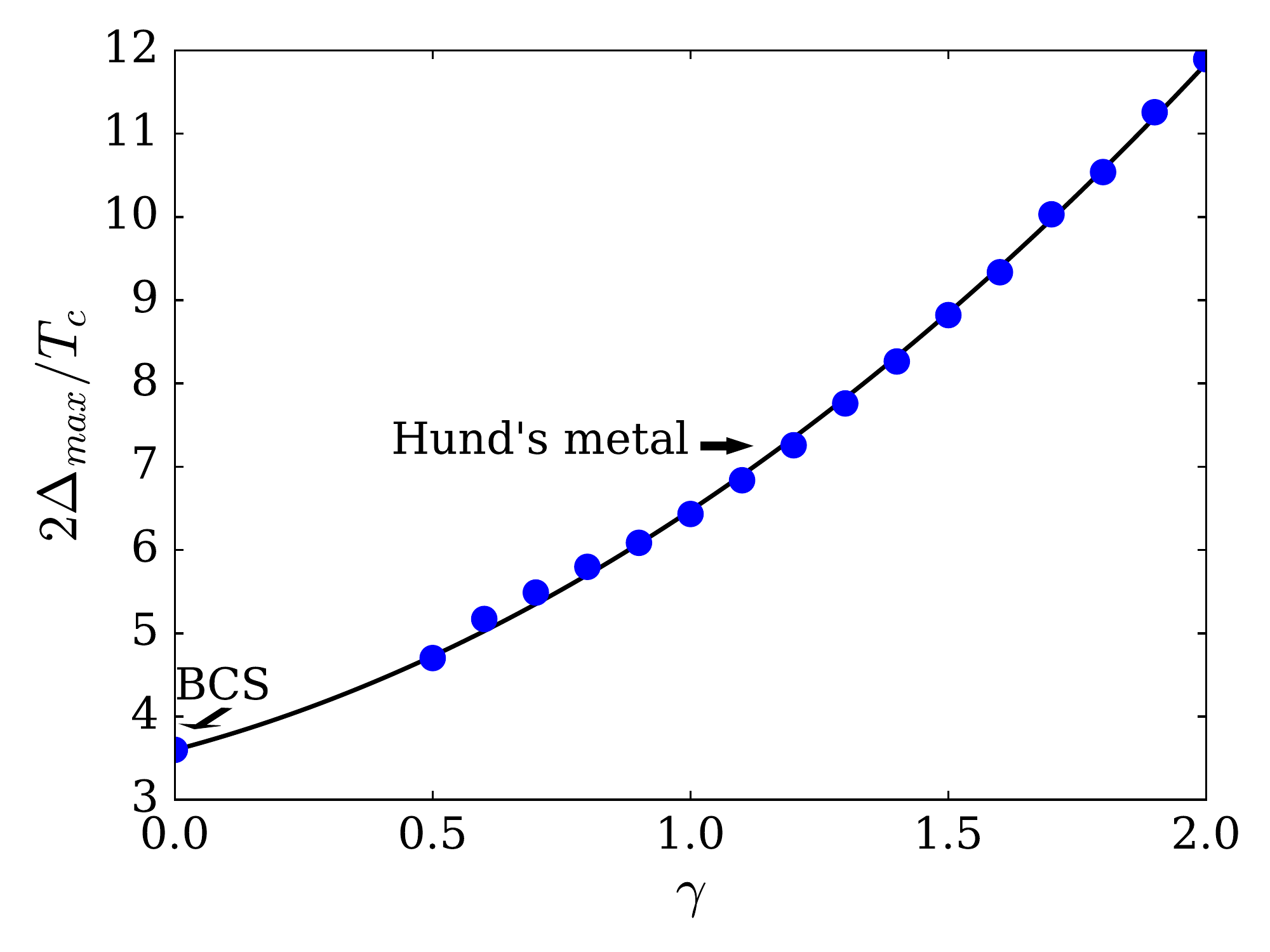}
\caption{(Color online) The ratio $2\Delta_{\max}/T_c$ as a function of $\gamma$. The arrows indicate the ratio $2\Delta_{\max}/T_c=7.2$ ($\gamma\sim 1.2$) and $3.6$ ($\gamma=0$) corresponding to the ratio for FeSC and BCS. The black solid line is the fit to a parabola with $2\Delta_{\max}/T_c=3.6+1.63\gamma+1.24\gamma^2$.}
\label{fig2}
\end{figure}

\noindent $T_c=A(\gamma)\Omega_0$ and $\Delta_{\max}=B(\gamma)\Omega_0$, i.e., $2\Delta_{\max}/T_c = 2B(\gamma)/A(\gamma)$.

It is instructive to
study how the ratio $2\Delta_{\max}/T_c$ varies with the exponent $\gamma$ because different $\gamma$ describe different pairing mechanisms. We show our
numerical results for $2\Delta_{\max}/T_c$ for various $\gamma$ in Fig. \ref{fig2}. The ratio, $
2\Delta_{\max}/T_c$, increases
with increasing $\gamma$
 in a parabolic fashion which can be extrapolated to the BCS ratio, $2\Delta_{\max}/T_c=3.6$, at $\gamma \geq 0$.  We found that
  the experimental  $2\Delta_{\max}/T_c \sim 7.2$, reported by Miao {\em et al.},
  is reproduced for $\gamma\sim 1.2$.
Remarkably, this  value of $\gamma$  coincides
 with the exponent of
the local spin
  susceptibility,  obtained from the extensive numerical analysis of Hund's metal state in the three-band Hubbard model~\cite{NRG_susceptibility,Yin_susceptibility}.
  This agreement is the strong argument that
   incoherent
   spin fluctuations, specific to a Hund's metal  state,  may indeed mediate superconductivity in FeSCs.

To get further insight into this issue, we now discuss how
 $\Delta_{\max}$, $T_c$, and also fermionic $\Sigma (\omega_m)$  individually vary with $\gamma$.  To get $T_c (\gamma)$  and self-energy near $T_c$, we follow \cite{Moon_gamma_model,Wang_gamma_model} and (a) solve for $T$ at which the
  linearized gap equation (the one with infinitesimally small $\Delta (\omega_m)$) has the solution, and (b) solve Eq. \ref{eq:Z} at $\Delta_n =0$.

\begin{figure}[h]
\includegraphics[width=8.2cm]{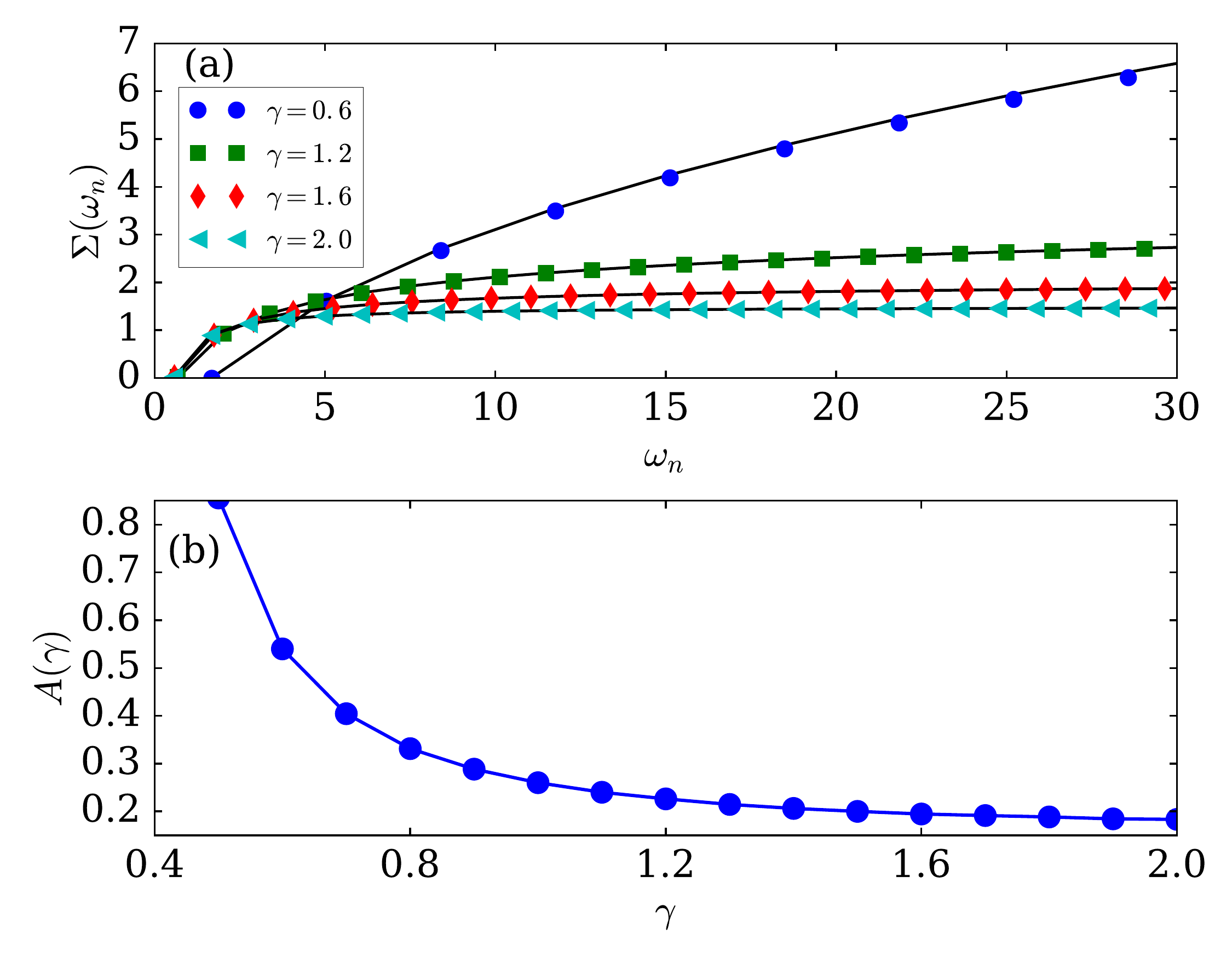}
\caption{(Color online) (a) The Matsubara self-energy $\Sigma(\omega_n)$ for different power $\gamma$, with $\Omega_0=1$ near critical temperature $T_c$. The black solid line is the analytic solution, from Moon and Chubukov \cite{Moon_gamma_model}. (b) The coefficient $A(\gamma)$ corresponding to the critical temperature $T_c=A(\gamma)\Omega_0$ as a function of power $\gamma$.}
\label{fig3}
\end{figure}

 We show  the result of numerical calculation of $\Sigma(\omega_n)$ in Fig. \ref{fig3}(a). Analytical reasoning shows~\cite{Moon_gamma_model} that, at
   large $\omega_n$, $\Sigma(\omega_m)$ scales as $\omega^{1-\gamma}$ for $0<\gamma <1$ and saturates to $\Sigma (\omega_m) = \big[\Omega^\gamma_0 /(2\pi T)^{\gamma -1}\big] \zeta (\gamma)$ for $\gamma >1$, where $\zeta(\gamma)$ is the Riemann zeta function. Our numerical results fully reproduce this asymptotic behavior.
In Fig. \ref{fig3}(b) we show the numerical result for the prefactor $A(\gamma)$ in the critical temperature,
  $T_c=A(\gamma)\Omega_0$.
   The analytical expression for $A(\gamma)$ has been obtained in Ref. \onlinecite{Wang_gamma_model}  within large $N$ approximation.  An extension of that result to 
    the physical case
  $N=1$ yields
    $A(\gamma)=\frac{1}{2\pi} \big(1+\frac{\delta_\gamma}{\gamma}\big)$,
   where $\delta_\gamma$ is a number in the order of one ($\delta_{\gamma \gg 1} \approx 1/2$). Our

\vspace{0.cm}

\noindent numerical result is consistent with this formula, particularly the increase of 
$A(\gamma)$ at smaller $\gamma$ and the  saturation of $A(\gamma)$ at $1/2\pi$ at larger $\gamma$ ( for $\gamma=2$ we found $A(2)=0.18$, in good agreement with Ref. \onlinecite{Wang_gamma_model}).

\begin{figure}[h]
\includegraphics[width=8.3cm]{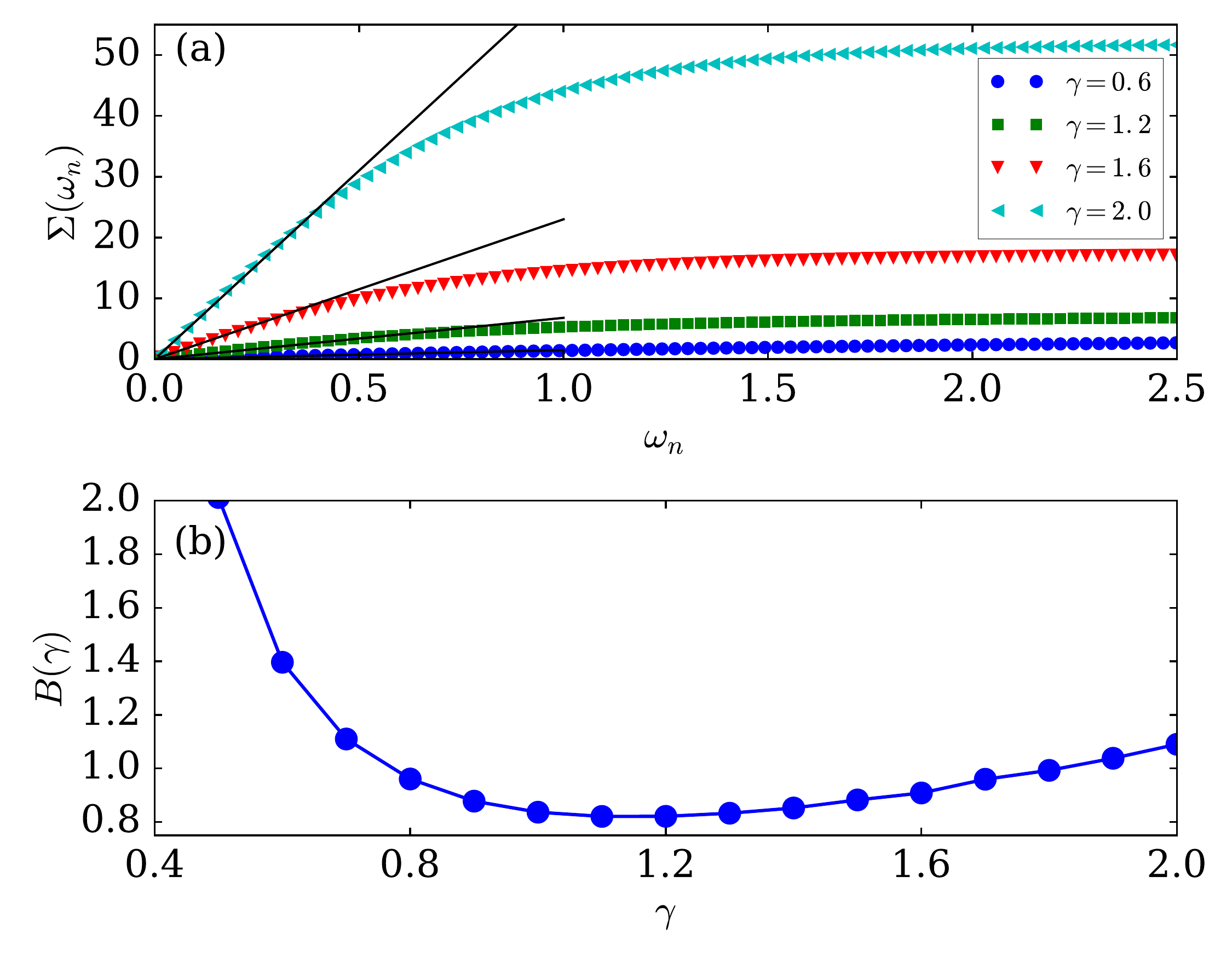}
\caption{(Color online) (a) The Matsubara self-energy $\Sigma(\omega_n)$ for different power $\gamma$ at $\Omega_0=1$ and $T=0.005\Omega_0$. The solid black lines indicate the low-frequency linear behavior. (b) The coefficient, $B(\gamma)$, corresponding to the maximum of the pairing gap $\Delta_{\max}=B(\gamma)\Omega_0$, as a function of the exponent $\gamma$. }
\label{fig4}
\end{figure}

At low temperatures, the pairing gap
$\Delta(\omega_n)$, is no longer a small quantity. The
linearization trick is no longer applicable, and we have to solve the full nonlinear gap equation for $\Delta(\omega_n)$ and convert the result to the real frequency axis.
In Fig. \ref{fig4} we show the results for the self-energy $\Sigma (\omega_n)$ and the prefactor $B(\gamma)$ in $\Delta_{\max}=B(\gamma)\Omega_0$, obtained from Eq. \ref{eq:Z} using the solution of $\Delta (\omega_n)$,  for  $T=0.005 \Omega_0$.
 The self-energy  $\Sigma (\omega)$ (Fig. \ref{fig4}(a)) scales linearly with $\omega_n$ at small frequencies, as expected in a Fermi liquid.  The restoration of Fermi-liquid behavior is the known feedback effect from superconductivity, which, e.g., accounts for peak-dip-hump behavior in cuprate superconductors below $T_c$ (see, e.g., Ref. [\onlinecite{Fink}]). In physical terms, this happens  because a finite gap reduces quasiparticle scattering at low frequencies and makes low-energy states longer-lived.
The slope of $\Sigma(\omega_n)$ at low frequency increases with increasing $\gamma$,   indicating that correlations get stronger.

The behavior of $B(\gamma)$ is shown in
 Fig. \ref{fig4}(b). At small $\gamma$,  $B(\gamma)$ decreases rather abruptly with increasing $\gamma$.  At larger $\gamma$, $B(\gamma)$ passes through a minimum at $\gamma\sim 1.2$ and slowly increases for $\gamma>1.2$.
 The ratio $2\Delta_{\max}/T_c$ (Fig. \ref{fig2}) is determined by the ratio between $B (\gamma)$ (Fig. \ref{fig4}(b)) and $A(\gamma)$ (Fig. \ref{fig3}(b)).
      At small $\gamma$, both $A(\gamma)$ and $B(\gamma)$  strongly evolve with $\gamma$, but the decrease of $B(\gamma)$ with increasing $\gamma$ roughly  follows the trend of $A(\gamma)$. As a result, the ratio $ 2B (\gamma)/A(\gamma)$ increases with increasing $\gamma$ but varies not as strongly as $A(\gamma)$ and $B(\gamma)$.
       For larger $\gamma >1.2$,
        $A(\gamma)$ saturates  and the enhancement of $2\Delta_{\max}/T_c$  is due to the increase of
         $B(\gamma)$.
        
\noindent

\begin{figure}[h]
\includegraphics[width=8.5 cm]{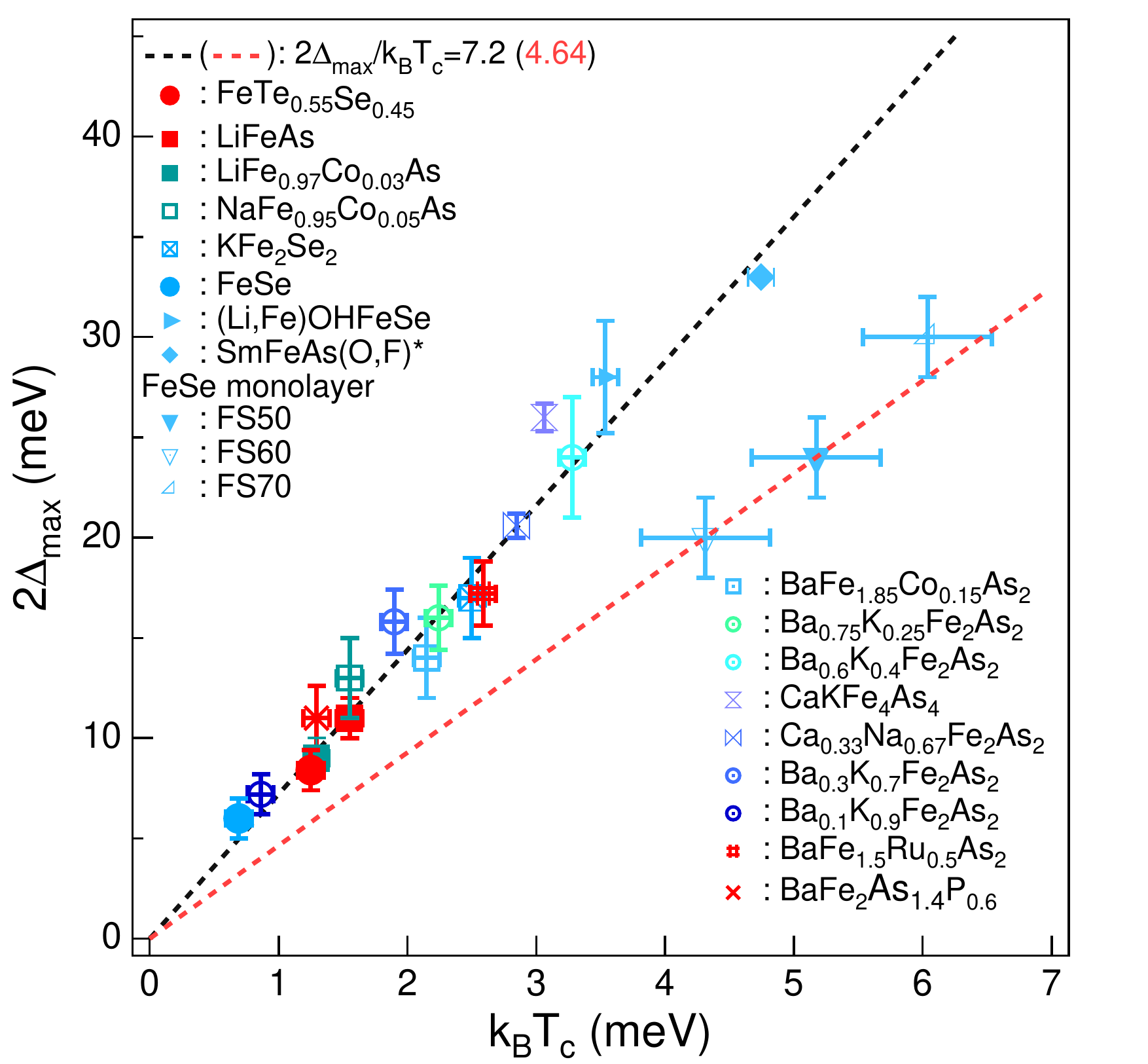}
\caption{(Color online) Summary of $2\Delta_{SC}^{\max}/k_{B}T_{c}$ that is determined by ARPES in various bulk FeSCs \cite{Terashima2009,Nakayama2011,WangXP2011,Xu2013,Liu2011,Miao2012,Miao2015,Miao2016,
Shi2014,Mou2016,Ding2008,Xu2013b,ZhangY2012,Liu2018} and monolayer FeSe films under different annealing conditions \cite{ Lee2014, Shi2017}. The black and red dashed lines are linear function fit of the bulk FeSCs and monolayer FeSe, respectively. Since SmFeAs(O,F) has a non-neutral cleaved surface, the value of $2\Delta_{SC}^{\max}/k_{B}T_{c}$ is extracted from the bulk sensitive optical conductivity measurement\cite{ Charnukha2018}. { Systematic errors 
due to the finite instrumental resolution and the profile of the superconducting peak give 10\%$\sim$17~\% uncertainty of the $2\Delta_{\max}/k_{B}T_{c}$ values.}}
\label{fig5}
\end{figure}

{\bf \it Discussion}~~~
Motivated by the theoretical understanding   we  reexamined  $ 2 \Delta_{\max} / T_c $ 
in other Fe-based superconductors. The data are summarized in Fig.  \ref{fig5}. We see that they fall onto a single curve with the same slope {$ 2 \Delta_{\max} / T_c = 7.2\pm1$} as in LiFeAs and FeTe$_{0.55}$Se$_{0.45}$. 
  This universality is the strong argument in favor of 
  Hund's metal  description with electronic 
  (
  spin fluctuation)
   pairing mechanism.   Remarkably,  the 
    data for FeSe monolayers 
     fall on a different curve with a smaller {$2 \Delta_{\max}/T_c = 4\pm0.5$}.
  This is consistent with the idea 
  that 
   in these systems the pairing may be mediated by   electron-phonon interaction~
    \cite{PhysRevB.97.060501,lee,1367-2630-17-7-073027,Zhang2017}.

FeSCs 
 are members of a broad class of unconventional superconductors, 
 which also include 
 copper
oxides, heavy fermion metals, and the  organic charge-transfer salts. 
 The  normal state of all these 
superconductors satisfies the criterion of bad metals, 
 superconductivity appears near an antiferromagnetic  phase, and 
$T_c$ is a sizable fraction of the bandwidth.
In this respect, FeSCs are often compared to the cuprates~\cite{Basov}, 
 because the bandwidths are comparable.  
  At a face value,  $2\Delta_{\max} / T_c $ in underdoped cuprates is larger. 
 However, one needs to take into account four additional considerations. First, the $d$-wave character of superconductivity in the cuprates modifies the
     $2\Delta_{\max}/T_c$ already in the BCS limit (Ref. [\onlinecite{betouras_karen}]).   Second, 
 superconductivity in underdoped cuprates emerges from a pseudogap regime, and  $\Delta_{\max}$ 
 (the gap in the antinodal region)
   develops at an energy scale $T^* > T_c$. It would then be more appropriate to relate it to $T^*$ rather than to $T_c$.  Third, even above optimal doping, when pseudogap effects are relatively weak, phase fluctuations are not negligible,  and the onset temperature $T_p$ for the emergence of the bound pairs is larger than $T_c$. Eliashberg theory neglects phase fluctuations and, within it, one can only get the $2\Delta_{\max}/T_p$ ratio.

\vspace{0.cm}
   
\noindent Fourth, there are substantial inhomogeneities in the sample, and one should  compare a local temperature $T_p$ and a local gap  $ \Delta_{\max} $  in a given region~\cite{Greven18}
   This has been done in the  tunneling studies~\cite{gomes}, which reported $2\Delta_{\max}/T_p\sim 7.9 $, not that far from FeSCs.
      If we take all this into consideration, 
       it appears that 
        FeSCs  and the 
         cuprates are  closer than one would have expected at first sight.

{ The heavy fermion superconductors also have similar $2\Delta_{\max}/T_c$ ratios. For example,
UPd$_2$Al$_3$ has $2\Delta_{\max}/T_c=6$ \cite{UPd2Al3_ratio} and PuCoGa$_5$ has $2\Delta_{\max}/T_c=6.4\pm0.4$ \cite{Pu_ratio}, which give $\gamma \sim 1.0$. Notice that in these systems the Hund's coupling is important. On the other hand, the organic charge-transfer superconductors have ratios $2\Delta_{\max}/T_c=4.8$ corresponding to $\gamma\sim0.5$ \cite{organic_ratio}, which are believed to be Mott systems and the Hund's physics is not relevant.}

Finally, we comment on  earlier realistic   calculations  of superconductivity in FeSCs.
Yin {\em et al.} investigated possible pairing states using  the
LDA+DMFT effective pairing interaction
 which describes the observed spectra\cite{Yin_pairing}. Because of the computational cost, they could  not
 go  to low enough temperatures to study $2\Delta_{\max}/T_c$ and/or  carry out an Eliashberg treatment. 
Nourafkan {\em et al.}
 solved the  Eliashberg
  equations  but replaced frequency-dependent interaction by a constant~\cite{Nourafkan_pairing}. { Ummarino carried out an Eliashberg treatment to FeScs and yields similar $2\Delta_{\max}/T_c$ ratios as ours \cite{Ummarino}. In his work,
the pairing interaction was introduced phenomenologically.}
Combining the realistic
 pairing interaction with the Eliashberg approach is an outstanding challenge for future work.

{\it Conclusions}~~~
In this work, we build on the
recent understanding of the physics of the Hund's metal
 and
  studied a phenomenological $\gamma$-model describing  the superconductivity mediated by bosonic propagator
   with a power-law frequency dependence,
    $\lambda(\Omega)\propto 1/|\Omega|^{\gamma}$.
 This model captures the essence of the transition from a Hund's metal to a
  superconductor
  at a temperature comparable to or higher  than a crossover temperature between non-Fermi-liquid and Fermi-liquid behavior
  ~\cite{NRG_susceptibility,Yin_susceptibility}.

We use the model to explore the main characteristics of the pairing gap and $T_c$, ignoring the complications
such as the
 multiorbital or multiband  structure of FeSCs.
 We find $2\Delta_{\max}/T_c$  to be independent of the interaction strength and equal to $7.2-7.3$ if we use $\gamma =1.2$ obtained from the three-band Hubbard model.
 These results are in surprisingly  good
   agreement
with
recent experiments
 which argued that $2\Delta_{\max}/T_c \approx 7.2$ is the same in at least two FeSCs: LiFeAs and FeTe$_{0.55}$Se$_{0.45}$ \cite{Miao_gap_Tc}.
It would be interesting to extend these observations to a more
realistic description of the materials, taking into account the
multiorbital nature of the problem, and the fact that, in Hund's metals, the
power-law
 behavior of local spin susceptibility holds in  an intermediate
  temperature range between a Fermi-liquid regime at low temperatures  and
a high temperature regime where the orbitals and the spins are both quasi-atomic-like.

\begin{acknowledgments}
We would like to thank  Ar. Abanov, K. Haule, K. Stadler, J. VonDelft, and Y. Wu for numerous discussions on the subject
of Hund's metals and superconductivity in the $\gamma$-model.  T.-H.L. and G.K. were supported by the NSF Grant No. DMR-1733071.  A.V.C. was supported by the NSF Grant No. DMR-1523036.  H. M. is supported by the U.S. Department of Energy, Office of Basic Energy Sciences, Early Career Award Program under Award No. 1047478.
\end{acknowledgments}

\bibliography{ref,supplemental}

\pagebreak
\widetext
\begin{center}
\textbf{\large Supplemental Materials: Pairing Mechanism in Hund’s Metal Superconductors and the Universality of the Superconducting Gap to Critical Temperature Ratio}
\end{center}
\setcounter{equation}{0}
\setcounter{figure}{0}
\setcounter{table}{0}
\setcounter{page}{1}
\makeatletter
\renewcommand{\theequation}{S\arabic{equation}}
\renewcommand{\thefigure}{S\arabic{figure}}

\section{Spin-Fluctuation Eliashberg Theory}
The Eliashberg equations after averaging over the Fermi surface reads \cite{spin-fluct-d-wave-book,Moon_gamma_model,Chubukov_colorSC},

\begin{equation}
Z(\omega_{n})=1+\pi T\sum_{m}\frac{\omega_{m}}{\omega_{n}}\lambda(\omega_{m}-\omega_{n})\frac{Z(\omega_{m})}{\sqrt{Z(\omega_{m})^{2}\omega_{m}^{2}+\Phi(\omega_{m})^{2}}},\label{eq:Eliashber_infD1_sup}
\end{equation}

\begin{equation}
\Phi(\omega_{n})=Z(\omega_{n})\Delta(\omega_{n})=\pi T\sum_{m}\lambda(\omega_{m}-\omega_{n})\frac{\Phi(\omega_{m})}{\sqrt{Z(\omega_{m})^{2}\omega_{m}^{2}+\Phi(\omega_{m})^{2}}},\label{eq:Eliashber_infD2_sup}
\end{equation}

\noindent where $Z(\omega_n)=1+\frac{\Sigma(\omega_n)}{\omega_n}$ is the quasiparticle renormalization factor, $\Delta(\omega_n)$ is the pairing gap, and $\Sigma(\omega)$ and $\Phi(\omega_n)$ is the normal and anomalous self-energy, respectively. The pairing interaction is defined as

\begin{equation}
\lambda(\Omega_{n})=gN(0)\chi'_{loc}(\Omega_n)=\int_{0}^{\infty}d\nu\frac{2g\nu N(0)}{\Omega_{n}^{2}-\nu^{2}}\chi''_{loc}(\nu),\label{eq:lambda_sup}
\end{equation}

\noindent where $\chi_{loc}(\omega_n-\omega_m)\equiv\int_{0}^{2k_F} dqq\chi(q,\omega_{n}-\omega_{m})$  is the local spin susceptibility \cite{Chubukov_colorSC} and we use the spectral representation of the susceptibility. 

Combining Eq. \ref{eq:Eliashber_infD1_sup} and Eq. \ref{eq:Eliashber_infD2_sup},
we are left with only one self-consistent equation,

\begin{equation}
\Delta(\omega_{n})=\pi T\sum_{\omega_{m}}\frac{\lambda(\omega_{m}-\omega_{n})}{\sqrt{\omega_{m}^{2}+\Delta^{2}(\omega_{m})}}\big(\Delta(\omega_{m})-\Delta(\omega_{n})\frac{\omega_{m}}{\omega_{n}}\big),\label{eq:Eliashberg_sup}
\end{equation}

\noindent and the quasiparitcle renormalization factor can be calculated
from

\begin{equation}
\omega_{n}Z(\omega_{n})=\omega_{n}+\pi T\sum_{\omega_{m}}\lambda(\omega_{m}-\omega_{n})\frac{\omega_{m}}{\sqrt{\omega_{m}^{2}+\Delta^{2}(\omega_{m})}}.\label{eq:Z_sup}
\end{equation}

\section{The power law-behavior in the spin-excitation of Hund's metals}

In this section, we show the power-law behavior in the spin-excitation of Hund's
metals based on the data in the Fig. 3(c) of Ref. \onlinecite{NRG_susceptibility} and our
continuous-time quantum Monte-Carlo (CTQMC) simulation, and
their relation to the pairing interaction, $\lambda(\Omega_n)$, as defined in Eq. \ref{eq:lambda_sup}.

\begin{figure}[h]
\includegraphics[width=7.6 cm]{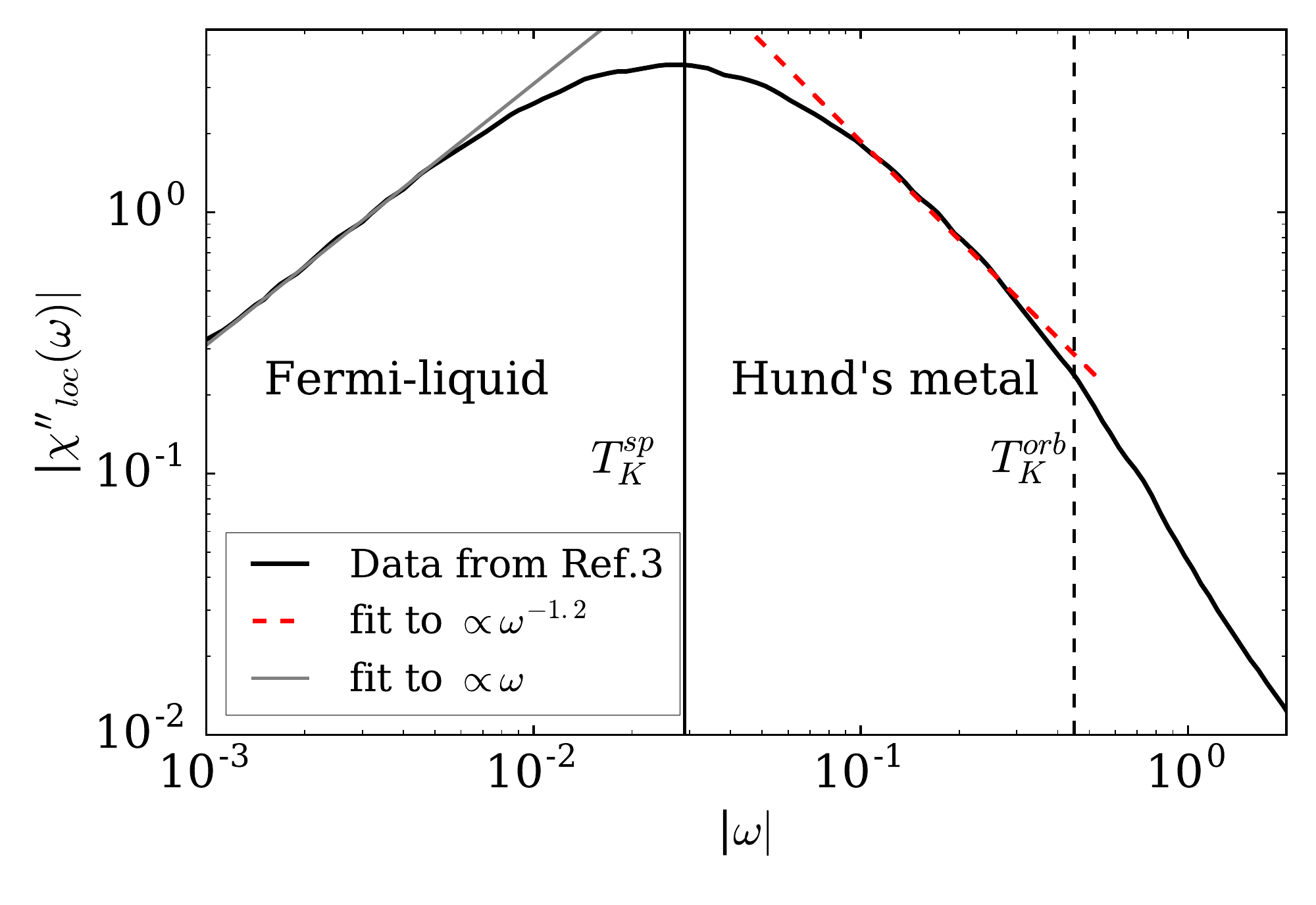}
\caption{The local spin susceptibility, $\chi''_{loc}(\omega)$, for three-band Hubbard model at $T=0.001t$ extracted from Fig. 3(c) of  Ref. \onlinecite{NRG_susceptibility}.}
\label{susc_fit}
\end{figure} 
 
Fig. \ref{susc_fit} shows the local spin susceptibility of the three-band Hubbard model, which
captures the essence of Hund's physics, extracted from the Fig. 3(c) of Ref. \onlinecite{NRG_susceptibility}. The vertical solid and dashed line in Fig. \ref{susc_fit}
denote the spin Kondo temperature, $T_K^{sp}$, and orbital Kondo temperature, $T_K^{orb}$,
respectively. Below $T_K^{sp}$, both the spin and orbital degrees of freedom are screened.
Therefore, the system shows a Fermi-liquid behavior characterized by a linear susceptibility,
$\chi''(\omega)\propto \omega$. In the intermediate regime, $T_K^{sp}<\omega<T_K^{orb}$,
the screened orbital degrees of freedom coupled to the slow fluctuating spins leading to
a fractional power-law behavior in the spin-excitation as indicated by the fit to
$\chi''(\omega)\propto \omega^{-1.2}$. The energy scale in this intermediate regime is pertinent to the superconducting temperature
of the iron-based superconductors. Therefore, the fractional spin-excitation may mediate the pairing
of the superconductivity. Having established the fractional power of the spin-excitation, it is straightforward to show from Eq. \ref{eq:lambda_sup} that the pairing interaction has the form $\lambda(\Omega_n)\propto |\Omega_n|^{-1.2}$ given that $\chi''(\omega)\propto sign(\omega)|\omega|^{-1.2}$.

\begin{figure}[h]
\includegraphics[width=7.6 cm]{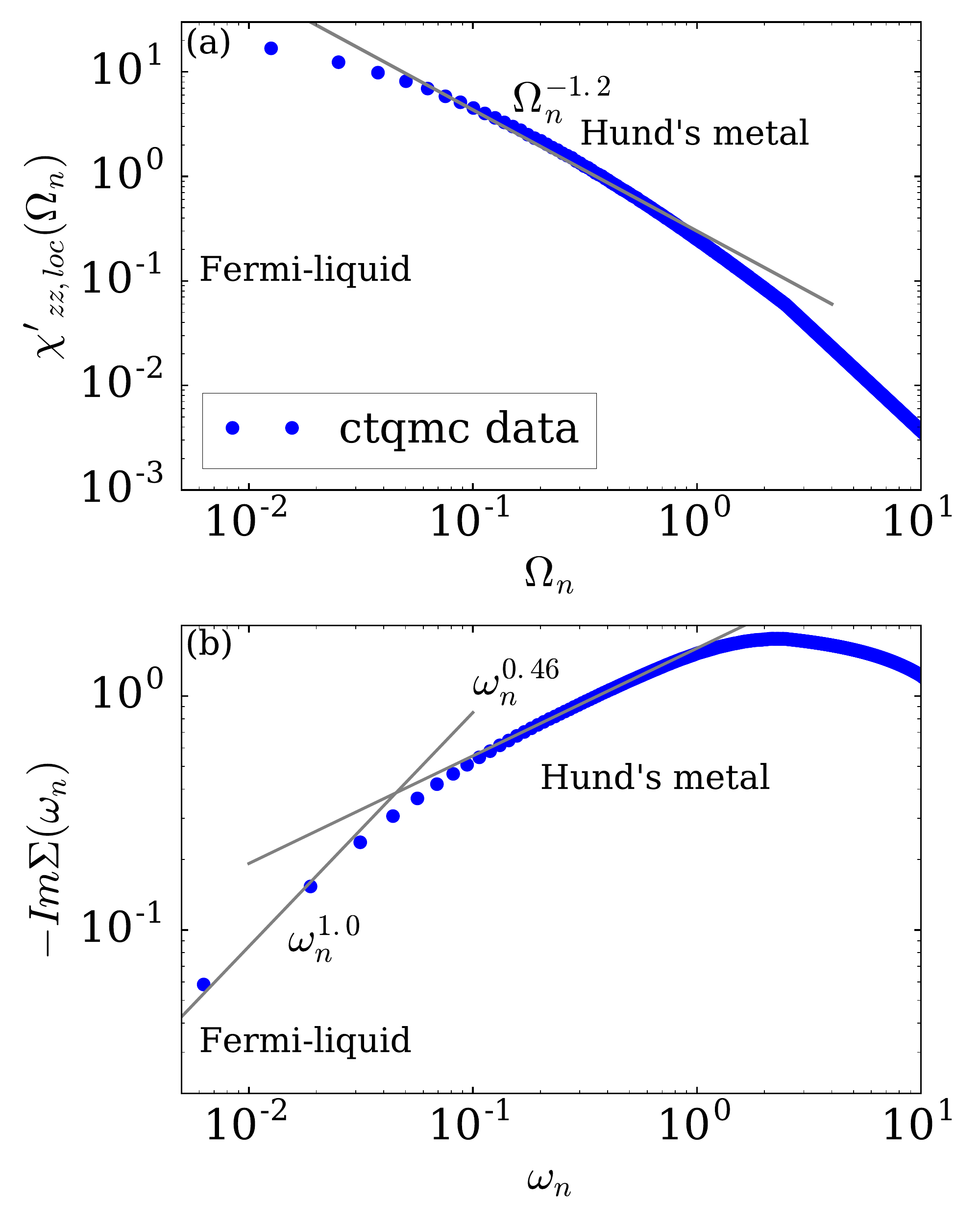}
\caption{(a) The z-component of the local spin susceptibility, $\chi'_{zz,loc}(\Omega_n)$, from CTQMC simulation on three-band Hubbard model at $T=0.002t$, $U=5t$, $J=1t$, and $\mu=7.032t$ where $t$ is the hopping amplitude (the same setting as in Fig. 1 of Ref. \onlinecite{NRG_susceptibility}). (b) The imaginary part of the self-energy, $Im\Sigma(\omega_n)$, with the same parameters.}
\label{susc_ctqmc}
\end{figure}

In Fig. \ref{susc_ctqmc}(a), we also present the z-component of the local spin susceptibility, $\chi'_{zz,loc}(\Omega_n)$, from our continuous-time quantum Monte Carlo (CTQMC) simulation on three-band Hubbard model, where we set the hopping, $t=1$, as our unit of energy, the Coulomb interaction, 
Hund's coupling interaction and the chemical potential are set to $U=5$, $J=1$, and $\mu=7.032$, 
respectively (the same setting as in Fig. 1 of Ref. \onlinecite{NRG_susceptibility}). The temperature, 
however, is set to $T=0.002t$, which corresponds to $T=10K$ if we set $t=0.5$ eV. The local spin susceptibility shows a power-law behavior, $\chi'_{zz,loc}(\Omega_n)\propto|\Omega_n|^{-1.2}$, in the intermediate energy scale and
saturate to a constant at low temperature. Note that the temperature is slightly higher than the one 
in Fig. \ref{susc_fit} so the Fermi liquid behavior (saturation to a constant susceptibility) is not 
fully established. From Eq. \ref{eq:lambda_sup}, it is straightforward to show that $\lambda(\Omega_n)\propto|\Omega_n|^{-1.2}$ given that $\chi'_{zz,loc}(\Omega_n)\propto|\Omega_n|^{-1.2}$, where we use the fact that $\chi'_{zz,loc}(\Omega_n)\approx\chi'_{loc}(\Omega_n)$ in our rotationally-invariant three-band Hubbard model. Figure \ref{susc_ctqmc}(b) shows the imaginary part of the self-energy, $Im\Sigma(\omega_n)$, with the same parameter setting as in Fig. \ref{susc_ctqmc}(a). A fractional power-law behavior, $Im\Sigma(\omega_n)\propto\omega^{0.46}$, also exist in the intermediate energy. At low temperature, the Fermi-liquid behavior, $Im\Sigma(\omega_n)\propto\omega$, is recovered.


\end{document}